\documentclass[
  journal=pasa,
  manuscript=research-paper, 
  year=2024,
  volume=37,
]{cup-journal}

\usepackage{microtype,siunitx,booktabs}
\usepackage{amsmath,amssymb}
\usepackage{hyperref}
\hypersetup{colorlinks,citecolor=blue,linkcolor=blue,urlcolor=blue}
\usepackage{aas-macros}

\title{Intranight optical monitoring of the rare quasar J0950+5128, the brightest known candidate for transition from radio-quiet to radio-loud state}

\author{Krishan Chand}
\affiliation{Department of Physics and Astronomical Science, Central University of Himachal Pradesh (CUHP), Dharamshala 176215, India}
\email[Krishan Chand]{krishanchand.kc007@gmail.com}

\author{Gopal-Krishna}
\affiliation{UM-DAE Centre for Excellence in Basic Sciences, Vidyanagari, Mumbai 400098, India}

\author{Hum Chand}
\affiliation{Department of Physics and Astronomical Science, Central University of Himachal Pradesh (CUHP), Dharamshala 176215, India}

\doi{10.1017/pasa.2024.32}

\received {dd Mmm YYYY}
\revised  {dd Mmm YYYY}
\accepted {dd Mmm YYYY}
\published{22 September 2024}


\begin{document}

\begin{abstract}
We report a novel pilot project to characterise intra-night optical variability (INOV) of an extremely rare type of quasar, which has recently been caught in the act of transiting from a radio-quiet to radio-loud state, on {a} decadal time scale. Such rare transitions may signify a recurrence, or conceivably the first switch-on of jet activity in optically luminous quasars. The newly formed jet could well be jittery and unsteady, both in power and direction. The optically brightest among such radio-state transition candidates, the quasar J0950+5128 ($z = 0.2142$), was monitored by us with dense sampling in the R-band, during 2020-21 in 6 sessions, each lasting $>$ 4 hours. This is the first attempt to characterise the INOV properties associated with this recently discovered, extremely rarely observed phenomenon of quasar radio-state transition. The non-detection of INOV in any of the 6 sessions, down to {the} 1-2\% level, amounts to {a} lack of evidence for a blazar-like optical activity, $\sim$ 2 years after its transition to radio-loud state was found. {The only INOV feature detected in J0950+5128 during our observational campaign was a $\sim$ 0.15-mag spike lasting < 6 minutes, seen at 13.97 UT on 18-March-2021.} We also report the available optical light curves of this quasar from the Zwicky Transient Facility (ZTF) survey, which indicate that it had experienced a phase of INOV activity around the time its transition to the radio-loud state was detected, however that phase did not sustain until the launch of our INOV campaign $\sim$ 2 years later.  
\end{abstract}

\section{Introduction}
\label{introduction}

Optical monitoring of many individual quasars, radio-loud and radio-quiet, spanning several decades has revealed that large intensity variations ( > 100\%) are exclusive to the former category. Prominent examples in this category, showing large outbursts (on month-like time scales), include the blazars  
AO0235+164: \citealp{Webb1989A&A...220...65W}; OJ 287: \citealp{Takalo1994VA.....38...77T}; CTA102: \citealp{Raiteri2017Natur.552..374R}; 1ES 1927+654: \citealp{Trakhtenbrot2019ApJ...883...94T}. Traditionally, this signature of strongly beamed jets of energetic particles has not been found associated with radio-quiet quasars (RQQs; e.g., \citealp{Padovani2017NatAs...1E.194P}), for which only a much milder optical variability is observed even on year-like time scales (e.g., \citealp{Macleod2012ApJ...753..106M}; \citealp{Caplar2017ApJ...834..111C}; \citealp{Sun2018ApJ...866...74S}; \citealp{Stone2022MNRAS.514..164S}; \citealp{Arevalo2023MNRAS.526.6078A}).
The persistent dichotomy in the degree of variability has led to the common perception that among quasars, any transition from {the} radio-quiet to radio-loud state does not happen on human time scales, 
although this almost certainly takes place on time scales of $\sim 10^{5-6}$ years (see, e.g., \citealp{Reynolds1997ApJ...487L.135R}; \citealp{Czerny2009ApJ...698..840C}; \citealp{An2012ApJ...760...77A}). Such variability time-scales are also expected by extrapolating from the compact stellar binaries, powered by $\sim 10 M_{\odot}$ black holes, for which jet-driven large radio variability is often witnessed on hour/day-like time scales (\citealp{Mirabel1999ARA&A..37..409M}).
However, over the past few years, exceptions to this popular perception about quasars have begun to surface, based on reports of detection of (first-time) switching-on, or restarting, of powerful jet activity in a few quasars, marked by between 6 to 25 fold jump in radio flux at centimetre wavelengths within a time span of at most 15-20 years
(\citealp{Mooley2016ApJ...818..105M}; \citealp{Kunert2020ApJ...897..128K}; \citealp{Nyland2020ApJ...905...74N}; \citealp{Wojtowicz2020ApJ...892..116W}; also, \citealp{Bannister2011MNRAS.412..634B}; \citealp{Bannister2011MNRAS.418.2813B}). {The} link of this jump in radio emission to newborn jets has been reinforced by the observed Gigahertz-Peaked-Specrtum (GPS) type (evolving) radio spectrum associated with such state transition episodes, as reported in the discovery papers cited above. Indications of changes occurring on such short time scales had earlier come from the targeted {Very Large Baseline Interferometry} (VLBI) campaigns which revealed nascent radio jets in quasars, with kinematical ages as small as $\sim$ 20 years (\citealp{Owsianik1998A&A...336L..37O}; \citealp{Owsianik1998A&A...337...69O}; \citealp{Gugliucci2005ApJ...622..136G}; \citealp{Orienti2021AN....342.1151O}). The detection of such young jets has bolstered the prospects of understanding the nature of the trigger for jet activity in {Active Galactic Nuclei} (AGN), which sometimes succeeds in pulling them out of the radio-quiet state. 

Arguably, the {best prospect of} AGN hosting the earliest cycles of jet activity is to be found in radio-loud narrow-line Seyfert1 (NLS1) galaxies (\citealp{Mathur2000MNRAS.314L..17M}). {The} existence of relativistic jets in such sources was indisputably established by their detection at gamma-rays (\citealp{Abdo2009ApJ...699..976A}), supported by independent observations, such as detection of high-amplitude intra-night optical variability (INOV) in several radio-loud NLS1 galaxies (e.g., \citealp{Liu2010ApJ...715L.113L}; \citealp{Paliya2013MNRAS.428.2450P}; \citealp{Kshama2017MNRAS.466.2679K}; \citealp{Gopal-Krishna2018BSRSL..87..281G}; \citealp{Paliya2019JApA...40...39P}; \citealp{Ojha2021MNRAS.501.4110O}). A very striking related clue has come from the discovery of strong, recurrent flares at millimetre wavelengths, on day-like time scales, in several NLS1s which are essentially radio-quiet at centimetre wavelengths (\citealp{Lahteen2018A&A...614L...1L})\footnote{The presence of relativistic jets in radio-quiet quasars, using millimetric observations, had earlier been inferred by \citet{Chini1989A&A...221L...3C}, using the IRAM telescope.}. These flares represent jumps in millimetre flux densities by factors of up to 10$^3$ in a matter of just a few days! The discovery paper poses the question whether this violent behaviour is {a} manifestation of an early-stage of AGN activity and whether its source, possibly a jittery and intermittent relativistic jet (see, e.g., \citealp{Czerny2009ApJ...698..840C}; \citealp{Lalakos2022ApJ...936L...5L}), is still confined within the {Broad Line Region} (BLR) region and, consequently, subject to free-free absorption, explaining the jet's faintness at centimetre/metre wavelengths (e.g., \citealp{Berton2020A&A...636A..64B}). This raises the possibility that a similarly violent activity on hour/day-like time scale might get triggered during the reported episodes of transition of quasars, from a radio-quiet to radio-loud state (see above). Here we attempt to test such a scenario by searching for strong intranight optical variability (INOV) which has now emerged as a reliable diagnostic of AGN activity driven by Doppler-boosted jet, including even the jets forming in low-mass AGN (e.g., \citealp{Gopal2023MNRAS.518L..13G} and references therein). The target chosen for our INOV monitoring campaign is the quasar J0950+5128 which is the optically brightest known candidate for radio-quiet to radio-loud state transition (\citealp{Nyland2020ApJ...905...74N}), out of the discovery of 26 such cases found by these authors to have undergone the radio state transition sometime during the last about two decades. Their radio emergence was unveiled during a comparison of the {Very Large Array} Sky Survey (VLASS, \citealp{Lacy2020PASP..132c5001L}) epoch 1 observations (2017-2019) with the existing FIRST (Faint Images of the Radio Sky at Twenty cm; 1993-2011) survey database (\citealp{Becker1995ApJ...450..559B}). The peak frequencies of the observed GPS spectra of these transition candidates, which have been measured for 14 of them using the {Very Large Array} (VLA), seem consistent with the very young (radio) ages of these quasars (\citealp{Nyland2020ApJ...905...74N}), as estimated using the radio size-peak frequency relation established for young AGN jets (\citealp{ODea1998PASP..110..493O}; \citealp{Jeyakumar2016MNRAS.458.3786J}).

As noted in \citet{Nyland2020ApJ...905...74N}, the $z = 0.2142$ quasar J0950+5128 (J095036.75+512838.12), which is powered by a black hole with an estimated mass of 10$^{8}$ $M_{\odot}$, was not detected in the FIRST survey (hence $<$ 0.44 mJy at 1.4 GHz) in April 1997. In April {2019}, however, its flux density was found to have risen to 8.77 mJy at 3 GHz, corresponding to a radio luminosity of 3.7$\times10^{40}$ erg/s. From VLA observations, its angular size was found to be $<$ 0.16 arcsec at 13.2 GHz (\citealp{Nyland2020ApJ...905...74N}). The SDSS catalogue gives its r-magnitude as 17.35, making it suitable for intranight monitoring with the 1.3-metre optical telescope accessible to us for this pilot experiment. The details of our optical observations, together with the derived differential light curves (DLCs) for 6 sessions are presented in Sect. \ref{Photometric monitoring}, followed by a brief discussion of the results in Sect. \ref{discussion} and the conclusions summarised in Sect. \ref{conclusions}.

\section{Photometric monitoring and data reduction} \label{Photometric monitoring}
\begin{figure*}
    \includegraphics[width=1.06\textwidth,height=1.0\textheight, trim=0.5cm 0cm 0.0cm 4.5cm,clip]{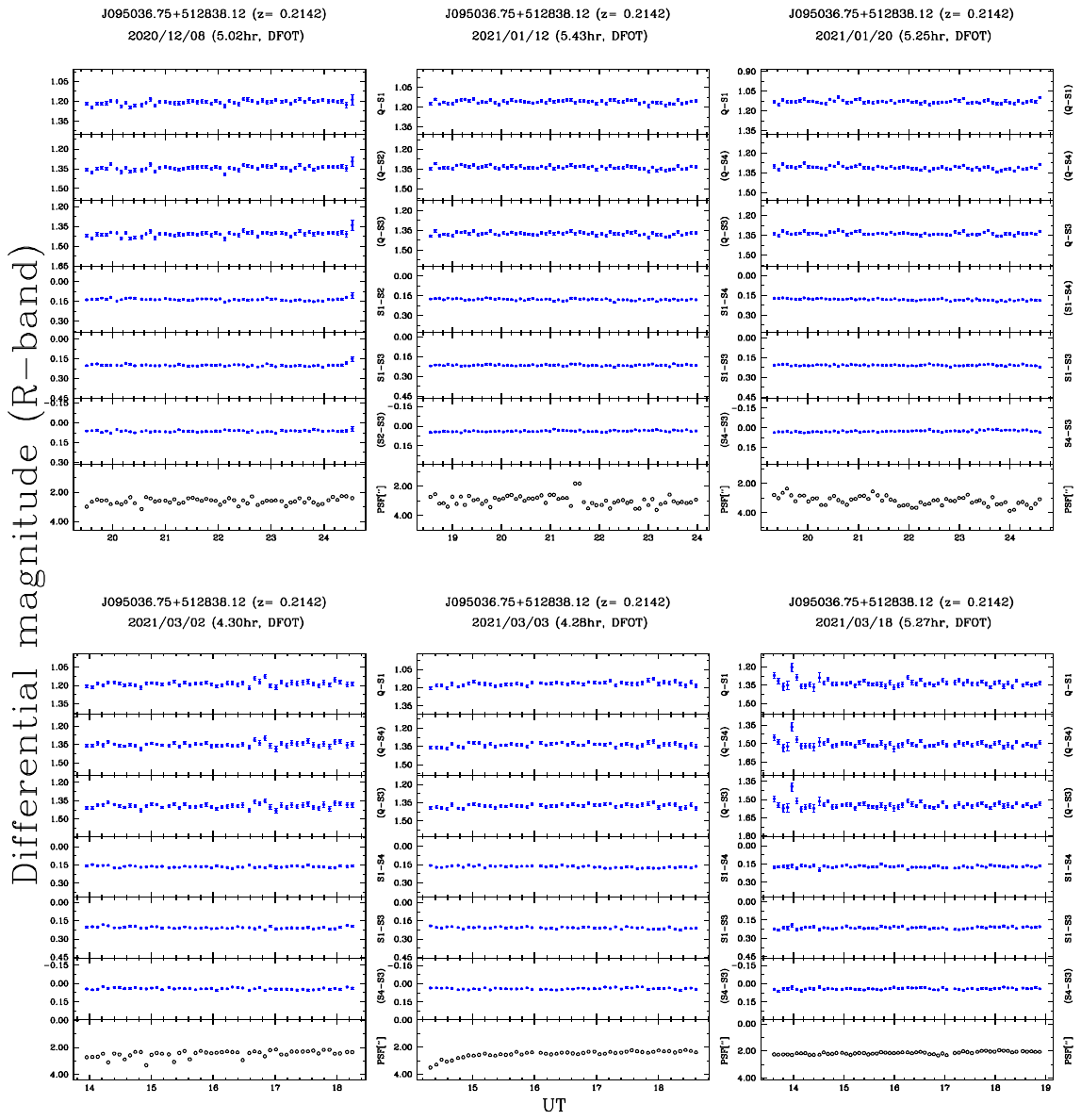}\\
  \vspace{-2.0in}
  \caption{Differential light curves (DLCs) for the six sessions. The name of the target object, together with the date and duration of its monitoring are shown near the top, in the mosaic for each session. The upper three panels in a mosaic display the `quasar-star' DLCs, while the lower three panels display the comparison `star-star' DLCs,
  as defined in the labels on the right side. The bottom panel for each session shows the seeing (PSF) variation during the session.}
\label{fig:all_dlc_part1} 
\end{figure*}

The (now radio-loud\footnote{Radio-loudness is classically defined as the ratio of the radio flux density at 5 GHz to the optical flux density at 4400 \AA, i.e., R = $\frac{f_{5 \, \mathrm{GHz}}}{f_{4400 \, \mathrm{\text{\AA}}}}$ \citep{Kellermann1989AJ.....98.1195K}. The quasars with R $\geq$ 10 are called radio-loud whereas those with R < 10 are called radio-quiet. For J0950+5128, in its radio-detected phase, we estimate R = 50.7, placing it securely in the radio-loud category.}) quasar J0950+5128 was observed in the Johnson–Cousins R band over 6 sessions using the 1.3-metre Devasthal Fast Optical Telescope (DFOT; \citealp{Sagar2011CSci..101.1020S}) at {Aryabhatta Research Institute of observational sciencES} (ARIES), near Nainital (India). In each session, the quasar was monitored continuously for a minimum duration of 4 hr, with each frame having a typical exposure of 4–5 min, recorded on 
a Peltier-cooled ANDOR CCD, equipped with 2k $\times$ 2k pixels (0.53 arcsec pixel$^{-1}$) and providing a field of view of 18.5 $\times$ 18.5 arcmin$^2$. The CCD detector operates with a gain of 2 e$^-$ per analogue-to-digital unit and exhibits a readout noise of 7.5 e$^-$ at a speed of 1 MHz. 
Pre-processing of the raw images involved bias subtraction, flat-fielding and cosmic-ray removal, all carried out using the standard tasks available in the Image Reduction and Analysis Facility (IRAF)\footnote{\url{http://iraf.noao.edu/}}. For each frame, the instrumental magnitudes of the quasar and the selected non-varying comparison stars within the same CCD frame were calculated through aperture photometry \citep[] {Stetson1987PASP...99..191S,1992ASPC...25..297S}, employing the Dominion Astronomical Observatory Photometry II (DAOPHOT II algorithm). The `point spread function' (PSF) for each frame was determined by averaging the full width at half maximum (FWHM) of the profiles of 5 moderately bright stars within the frame. {A} median of the PSF values found for all the frames in a session was taken as the `seeing' (FWHM of the PSF) for that session. For photometry, an aperture radius of 2$\times$FWHM (median) was adopted for that session (see, e.g., \citealp{Gopal2023MNRAS.518L..13G}, also, \citealp{Vibhore2023MNRAS.524L..66N}). The lower panel of Fig. \ref{fig:all_dlc_part1} displays the PSF variation during the session.
Differential light curves (DLCs) for each session were then generated for all the pairs involving the target quasar and the selected three steady comparison stars and these are displayed in Fig. \ref{fig:all_dlc_part1}.

\begin{table*}
\centering
\begin{minipage}{180mm}
\caption{Basic parameters of the quasar and the selected comparison stars for the 6 monitoring sessions.
\label{tab:comparison_color}}
\resizebox{\textwidth}{!}{
\begin{tabular}{ccc ccc c}\\
\hline

{Object  } &   Date of monitoring      &   {R.A.(J2000)} & {Dec.(J2000)}                      & {\it g} & {\it r} & {\it g-r} \\
           &  yyyy/mm/dd    &   (hh:mm:ss)       &($^\circ$: $^\prime$: $^{\prime\prime}$)   & (mag)   & (mag)   & (mag)     \\
{(1)}      & {(2)}        & {(3)}           & {(4)}                             & {(5)}   & {(6)}   & {(7)}     \\
\hline\\
J0950$+$5128 & 2020/12/08, 2021/01/12, 2021/01/20, 2021/03/02, 2021/03/03, 2021/03/18  &09:50:36.75 &$+$51:28:38.12 &18.11  &17.35   &0.76   \\
  S1         & 2020/12/08, 2021/01/12, 2021/01/20* ,2021/03/02, 2021/03/03, 2021/03/18            &09:49:42.95 &$+$51:28:13.79 &17.68  &16.30   &1.38  \\
  S2         & 2020/12/08*                                                                   &09:49:57.22 &$+$51:34:15.66 &17.18  &16.12   &1.06  \\
  S3         & 2020/12/08*, 2021/01/12*, 2021/01/20, 2021/03/02*, 2021/03/03*, 2021/03/18*            &09:49:37.18 &$+$51:36:00.32 &16.59  &16.00   &0.59  \\
  S4         & 2021/01/12*, 2021/01/20*, 2021/03/02*, 2021/03/03*, 2021/03/18*            &09:50:49.83 &$+$51:27:15.26 &16.96  &16.04   &0.92  \\\\
  \hline
\multicolumn{7}{l}{The comparison star shown in column 1 has been used for testing the presence of INOV of the quasar on the date marked with an asterisk.}\\
\end{tabular}
}
 \end{minipage}
\end{table*}

\begin{figure*}
\includegraphics[width=1.05\textwidth, height=0.525\textheight, trim=0.7cm 0cm 0.0cm 0.0cm,clip]{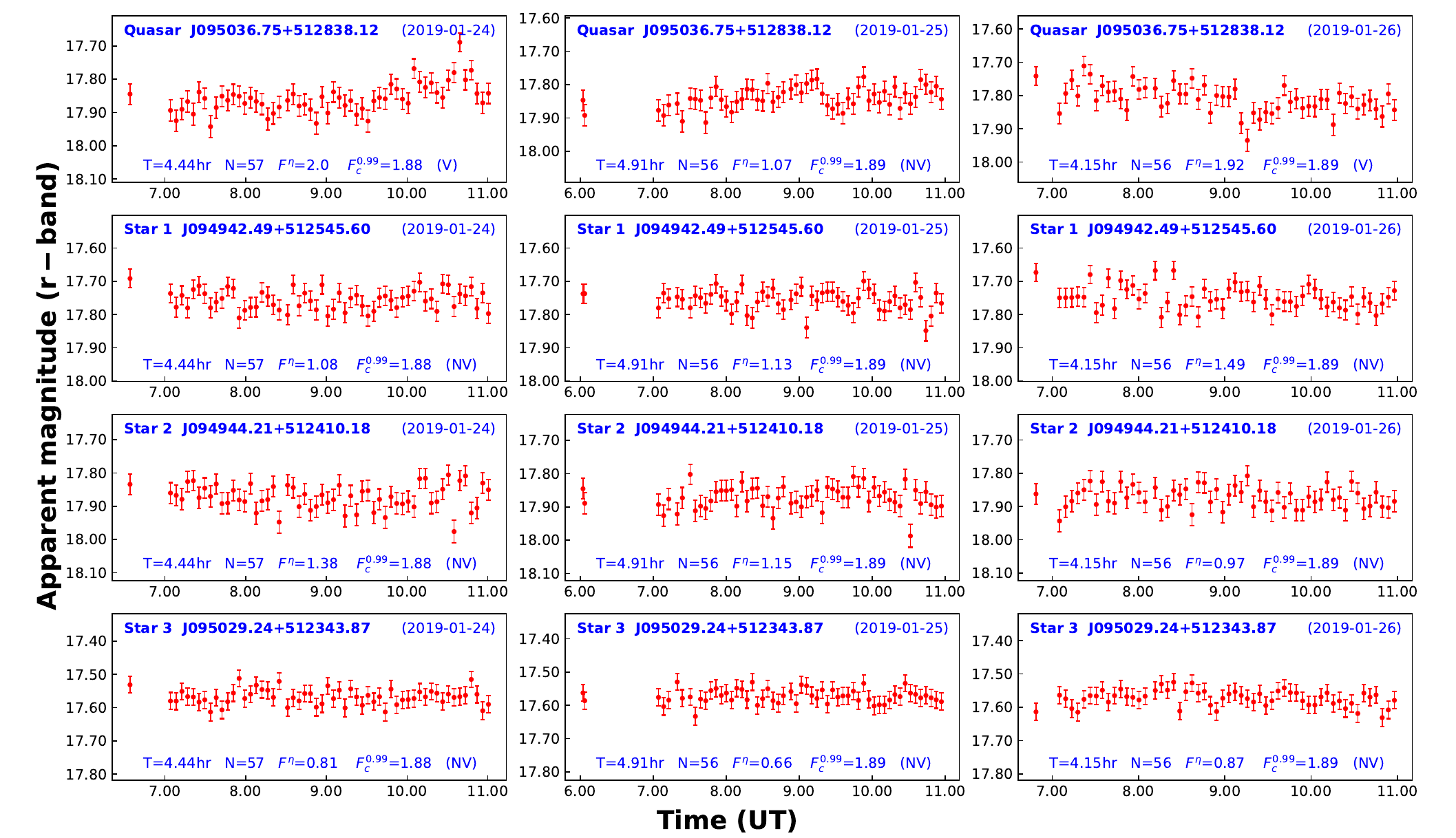}\\
  \vspace{-0.1in}
  \caption{The ZTF (r-band) light curves (LCs) of the quasar J0950+5128, for the three sessions. For each session, the top panel displays the LC of the quasar, while the lower 3 panels show the LCs for three selected comparison stars which are close in brightness to the target quasar. Each panel shows the SDSS name of the target (quasar/star) and the date and duration of ZTF monitoring. The INOV parameters determined from the variability analysis are shown below the corresponding LC, for each session (see text).}
  \label{fig:all_dlc_part2} 
\end{figure*}

\section{Statistical Analysis}
\label{stat}
To assess the presence of INOV in the DLCs, the widely used $F_{\eta}$ test \citep{Diego2010AJ....139.1269D} was applied, following the basic procedure described in \citet{Sapna2019MNRAS.489L..42M}, \citet{Gopal2023MNRAS.518L..13G} and \citet{Vibhore2023MNRAS.524L..66N}. For each session, the three comparison stars were initially selected (Table \ref{tab:comparison_color}) and the steadier two of them were identified by inspecting the `star–star' DLCs. The $F_{\eta}$ test was then applied to only the three DLCs involving these two comparison stars and the target quasar (see also, \citealp{Krishan2022MNRAS.511L..13C}). The selected pair of comparison stars used for each session is indicated within parentheses, in column 5 of Table \ref{tab2}. Additionally, the labels of the three DLCs that involve these two comparison stars and the target quasar, are shown within parentheses on the right-hand side in Fig. \ref{fig:all_dlc_part1}.\par
It has been found in several independent studies that the photometric errors returned by DAOPHOT are too small, by a factor $\eta$ ranging between 1.30 and 1.75 {and therefore ignoring this factor can sometimes lead to {the} spurious claim of INOV} \citep{Gopal-Krishna1995MNRAS.274..701G,Garcia1999MNRAS.309..803G,Sagar2004MNRAS.348..176S,Stalin2004MNRAS.350..175S,Bachev2005MNRAS.358..774B,Goyal2012A&A...544A..37G,Goyal2013MNRAS.435.1300G}. An accurate estimate of $\eta$ $ = 1.54 \pm 0.05$ was made through an analysis employing a large set of 262 DLCs of pairs of steady comparison stars. These stars had been monitored in 262 intranight sessions of {a} minimum 3 hr duration, together with the respective target AGN \citep{Goyal2013JApA...34..273G}. This accurate estimate of $\eta  = 1.54$ has been adopted in the present study.
The $F$-values calculated for the selected two quasar DLCs during each session are:
\begin{equation} 
\label{eq.ftest2}
F_{1}^{\eta} = \frac{Var(q-s1)}
{ \eta^2 \sum_\mathbf{i=1}^{N}\sigma^2_{i,err}(q-s1)/N},  \\
\hspace{0.1cm} F_{2}^{\eta} = \frac{Var(q-s2)}
{ \eta^2 \sum_\mathbf{i=1}^{N}\sigma^2_{i,err}(q-s2)/N}  \\
\end{equation}\\
where $Var(q-s1)$ and $Var(q-s2)$ are the variances of the DLCs of the target quasar, relative to the selected two comparison stars,
and $\sigma_{i,err}(q-s1)$ and $\sigma_{i,err}(q-s2)$ denote the rms errors returned by DAOPHOT on the $i^{th}$ data point in the DLCs of the target quasar, relative to the two comparison stars.
$N$ denotes the number of data points in the DLCs of the session (Column 3 of Table \ref{tab2}) and the scaling factor $\eta = 1.54$, as mentioned above. For each monitoring session, Table \ref{tab2} provides the values of $F_{1}^{\eta}$ and $F_{2}^{\eta}$ computed for the DLCs of the target quasar, relative to the two selected comparison stars. Also shown are the computed critical values of $F $ ($= F_{c}^{\alpha}$), taking $\alpha = $ 0.05 and 0.01 which correspond to the INOV detection confidence levels of 95\% and 99\%, respectively (also listed in columns 6 and 7 of Table \ref{tab2}, for each session). These critical values are meant for comparison with the $F-$values computed for the two DLCs of the quasar, using Eq. (\ref{eq.ftest2}), namely, $F_{1,2}^{\eta}$ and presented in column 5 of the Table \ref{tab2}. 
The target quasar is classified as `variable' (V) in a session if its computed $F-$value is $\ge$ $F_{c}$(0.99). If the $F-$value falls between $F_{c}$(0.95) and $F_{c}$(0.99), the DLC is considered as a `probable variable' (PV). In case the $F-$value is found to be less than $F_{c}$(0.95), the DLC is designated as `non-variable' (NV). Column 10 of Table \ref{tab2} lists the `Photometric Noise Parameter' for each session, defined as PNP = {$\sqrt { \eta^2\langle \sigma^2_{i,err} \rangle}$ }, where $\eta=1.54$, as mentioned above.

\begin{table*}
\centering
 \begin{minipage}{180mm}
\caption{Result of the statistical test for the presence of INOV in the DLCs of the Quasar J0950+5128.}
\label{tab2}
\resizebox{\textwidth}{!}{%
\begin{tabular}{cccccccccccc}
  \hline\\
\multicolumn{1}{c}{Quasar}
&\multicolumn{1}{c}{Date of obs.}
&{N}
&{T} 
&\multicolumn{3}{c}{F-test}
 &\multicolumn{2}{c}{INOV detection}
 
 &\multicolumn{1}{c}{$\sqrt { \eta^2\langle \sigma^2_{i,err} \rangle}$}
 & \\
  (SDSS name)       & yyyy/mm/dd & &{(hr)} &{$F_{1}^{\eta}$},{$F_{2}^{\eta}$}&{$F_{c}(0.95)$}&{$F_{c}(0.99)$}&\multicolumn{2}{c}{status$^{a}$} && \\
 (1)     &   (2)    &(3)&(4)  & (5)                         &(6)       &(7)           &(8)            &(9)     &(10)     &     \\\\
\hline\\
J095036.75$+$512838.12 &2020/12/08 &57&5.02	&  0.70 (S2),   0.71 (S3)&  1.56&  1.88& NV (S2)     ,     NV (S3) & NV    &    0.009 (S2-S3) &  \\
J095036.75$+$512838.12 &2021/01/12 &63&5.43	&  0.67 (S4),   0.63 (S3)&  1.52&  1.82& NV (S4)    ,      NV (S3)   &NV &    0.006 (S4-S3)&   \\                                                                                        
 J095036.75$+$512838.12 &2021/01/20 &63&5.25	& 0.71 (S1),    0.95 (S4)&  1.52&  1.82& NV (S1)     ,      NV (S4) & NV   &    0.007 (S1-S4)&  \\
 J095036.75$+$512838.12 &2021/03/02 &50&4.30	& 0.56 (S4),    0.59 (S3)&  1.61&  1.96& NV (S4)    ,      NV (S3)   & NV&    0.009 (S4-S3)&\\
 J095036.75$+$512838.12 &2021/03/03 &50&4.28	& 0.62 (S4),    0.48 (S3)&  1.61&  1.96& NV (S4)    ,      NV (S3)   & NV&    0.008 (S4-S3)& \\
 J095036.75$+$512838.12 &2021/03/18 &60&5.27	& 0.86 (S4),    0.94 (S3)&  1.54&  1.85& NV (S4)    ,      NV (S3)   & NV&    0.010 (S4-S3)&\\\\

\hline
\multicolumn{11}{l}{$^a$ V=variable, i.e., confidence level $\ge 0.99$; PV = probable variable ($0.95-0.99)$; NV = non-variable ($< 0.95$).}\\
\multicolumn{11}{l}{Variability status identifiers (col. 8), estimated for the `quasar- star1' and `quasar - star2' DLCs are separated by a comma.}
\end{tabular}
}
 \end{minipage}
\end{table*}

\section{Discussion}
\label{discussion}
{In this section, we shall briefly discuss the results of our search for INOV in the radio-state transition quasar J0950+5128.}

\subsection{The intranight monitoring with DFOT}

From Table \ref{tab2} it is seen that the quasar DLCs for all six sessions are consistent with {the} non-detection of INOV. {For blazars, a duty cycle of INOV detection has been found to be close to $50\%$, based on a large sample subjected to observations and analysis procedures very similar to that adopted in this work (\citealp{Goyal2013MNRAS.435.1300G}). Therefore, the non-detection of INOV on all six nights is very unlikely to be a chance occurrence.} Thus, there is little evidence for blazar-like activity in the wake of the quasar's change from a radio-quiet to radio-loud state, as found by \citet{Nyland2020ApJ...905...74N}. As mentioned in Sect. \ref{introduction}, our expectation of blazar-like INOV following the radio state-transition stems from the extreme-flaring events detected in the millimetric observations of a few NLS1 galaxies, on day-like time scales, despite their showing a highly muted variability at centimetre wavelengths (\citealp{Lahteen2018A&A...614L...1L}).  The contrast could
either be intrinsic, or due to absorption, or alternatively due to the non-simultaneity of the available data in the two bands (\citealp{Jarvela2024MNRAS.532.3069J}). In any case, it is expected that, compared to the muted variability at centimetre wavelengths, 
the (violent) variability seen at millimetre wavelengths would be more closely reflected in the optical band. This, however, is clearly not seen at least in the 6 monitoring sessions of this radio-state transition candidate quasar. 
Conceivably the INOV non-detection could be because of a misalignment of the jet of this quasar with respect to the line-of-sight, which would then call for an extension of the INOV campaign to other state transition candidate quasars. Moreover, in view of the growing evidence using the VLBI data archives, that the nuclear jets can undergo large directional changes on human time scales (e.g., \citealp{Britzen2023Univ....9..115B}; \citealp{Britzen2023ApJ...951..106B})\footnote{Note, however, that the outcome of such changes in jet orientation is circumscribed by the observed persistence of blazar activity in quasars over at least several decades (\citealp{Krishan2022MNRAS.516L..18C}; \citealp{Krishan2023PASA...40....6C}).}, an INOV search would be worthwhile in future observations of this quasar and, possibly, also in any existing independent datasets available for this and other radio state-transition quasars.

\subsection{Intranight light curves from the ZTF survey}

For the present quasar J0950+5128, which is currently the brightest known such case, such an opportunity is afforded by the Zwicky Transient Facility (ZTF) survey\footnote{\url{https://www.ztf.caltech.edu}}. A check of the ZTF database, applying the criterion of > 25 data points per session (of minimum 3.5 hr duration), has revealed that such intranight monitoring data are available for 3 nights for the present quasar, all taken during January 2019, i.e., just a few months before its transition to radio-loud state was recorded in the VLASS observations
(Sect. \ref{introduction}). Those 3 ZTF light curves (LCs) are reproduced in Fig. \ref{fig:all_dlc_part2}, together with the results of the statistical test we performed to check for the presence of INOV, again following the method described in Sect. \ref{stat} (note that the rms error on each photometric data point was taken directly from the ZTF database and so the parameter $\eta$ was set equal to unity). From the quantitative results of the $F_{\eta}$ test, displayed in Fig. \ref{fig:all_dlc_part2}, it is seen that statistically significant INOV was present during two out of the available three ZTF sessions for this quasar, all during January 2019. Also shown in the same figure are the ZTF LCs of three (simultaneously monitored) comparison stars, which match the quasar to within 0.25-mag (r-band). Neither star showed a statistically significant INOV in any of the 3 sessions, suggesting that the inferred statistical significance of the INOV of the quasar in two of the three ZTF sessions (January 2019) is probably real. This raises the possibility that the present quasar J0950+5128 was going through an INOV phase around the time its transition from a radio-quiet to radio-loud state was discovered (in April 2019, Sect. \ref{introduction}). However, its optical {variability} had subsided by the time our INOV campaign began in December 2020 (Table \ref{tab2}). 

A sustained follow-up of this pilot experiment is warranted, in order to develop a proper understanding of the relationship between the two extraordinary phenomena, namely the radio-quiet to radio-loud state transition of quasars and the early blazar-like activity hinted by the observations of violent millimetric flaring of (possibly nascent) AGN (Sect. \ref{introduction}). Such a possibility {of activity}, although hinted in the present study, is still of a preliminary nature, but a follow-up holds considerable promise of unravelling any underlying pattern. 

\subsection{A single-point brightness spike}

We conclude the present discussion by drawing attention to the only INOV feature detected in the DLCs of J0950+5128 during our campaign reported here. The feature is a $\sim$ 0.15-mag spike lasting $<$ 6 minutes, which was seen at 13.97 UT on 18-March-2021, in all 3 DLCs of the quasar taken relative to the 3 different comparison stars (see Fig. \ref{fig:all_dlc_part1}). For this reason, and also because the seeing disk had remained steady at that time (Fig. \ref{fig:all_dlc_part1}), we {conclude} that the sharp spike is probably real. It is interesting to recall that 15 highly significant single-point spikes of duration $<$ 5-10 minutes were identified in the DLCs (\citealp{GopalKrishna2000MNRAS.314..815G}; \citealp{Stalin2004MNRAS.350..175S}) under the AGN INOV program conducted at ARIES, which typically allowed INOV detection down to {the} $1 - 2\%$ level (see, \citealp{Gopal-Krishna2018BSRSL..87..281G}). Interestingly, 6 of the strongest 7 spikes (out of the total {of} 15), were observed in the DLCs of radio-quiet (or radio lobe dominated) quasars, and not in the {concurrently recorded} DLCs of the comparison stars. This hints at the possibility of an AGN related (albeit, a non-blazar) origin of the strong optical spikes.
With the rapid growth in optical databases on AGN monitoring, it would be interesting to look for similar spikes in their optical light curves, including those of the rare quasars found to have recently transitted from a radio-quiet to radio-loud state, or vice versa.  

\section{Conclusions}
\label{conclusions}

In this work, we have endeavoured to address the question {of} whether the transition from a radio-quiet to radio-loud state on {a} decadal time scale, as noticed a few years ago for a handful of quasars, is accompanied by some change in their activity level in the optical band. Thus, for the quasar J0950+5128, which is the optically brightest such transition case, we have enquired if post-transition it showed signs of a blazar-like activity. This possibility is hinted {at}, e.g.,  by the intense flaring of millimetric emission on a day-like time scale, witnessed in a few narrow-line Seyfert1 galaxies (which have been proposed to represent the early stage in the formation of radio-loud quasars). {Since} intranight optical variability (INOV) {is} a robust marker of blazar activity, we have carried out sensitive and densely sampled intranight optical monitoring of J0950+5128, on 6 nights between December 2020 and April 2021. Down to the 1-2\% detection limit, we found no evidence for INOV on any of the 6 nights, and hence our monitoring campaign has not unveiled any INOV activity in this quasar (except for a single $\sim$ 0.15-mag spike lasting < 6 minutes, observed on the night of 18-March-2021). Additionally, we find that for 3 nights, a few months before its radio-state transition was picked in the VLASS survey at 3 GHz, densely sampled ZTF r-band light curves are available for this quasar (during January 2019) and these show statistically significant INOV on two of the nights. This would suggest that around the time its transition to a radio-loud state was first noticed, the quasar exhibited some INOV activity which, however, subsided over the next two years, i.e., by the time of our INOV campaign. This, admittedly marginal evidence for a level change in the INOV activity, could either have an origin intrinsic to the jet, or merely reflect an increase in the jet's misalignment to the line-of-sight. The observational pointers presented here call for a follow-up of this first-time INOV study of a quasar which exemplifies an exceedingly rare phenomenon of radio-state transition in quasars occurring on {a} decadal, or possibly even shorter time scale.

\section*{Acknowledgments}
KC acknowledges the support of SERB-DST, New Delhi, for funding under the National Post-Doctoral Fellowship Scheme through grant no. PDF/2023/004071. G-K acknowledges a Senior Scientist position awarded by the Indian National Science Academy (INSA). The assistance from the scientific and technical staff of ARIES DFOT is thankfully acknowledged. The local hospitality extended to KC by IUCAA, Pune, during the preparation of this manuscript, and IUCAA associateship to HC are also thankfully acknowledged. This work is also based on observations obtained with the Samuel Oschin Telescope 48-inch and the 60-inch Telescope at the Palomar
Observatory as part of the Zwicky Transient Facility project. ZTF is supported by the National Science Foundation under Grants No. AST-1440341 and AST-2034437 and a collaboration including current partners Caltech, IPAC, the Oskar Klein Center at Stockholm University, the University of Maryland, University of California, Berkeley, the University of Wisconsin at Milwaukee, University of Warwick, Ruhr University, Cornell University, Northwestern University and Drexel University. Operations are conducted by COO, IPAC, and UW.
\section*{Data availability}
The data used in this study will be shared at a reasonable request by the corresponding author.




\bibliography{references}



\end{document}